\documentstyle[12pt,aasms4,epsf]{article}

\newcommand\teff{\ifmmode T_{\mathrm eff} \else $T_{\mathrm eff}$\fi}

\begin{document}

\title{The MACHO Project Large Magellanic Cloud Variable Star Inventory. VIII.
The Recent Star Formation History of the LMC from the Cepheid Period Distribution}

\author{C. Alcock}
\affil{Lawrence Livermore National Laboratory, Livermore, CA 94550 \\
E-mail: {\tt alcock@beowulf.llnl.gov}}

\author{R.A. Allsman}
\affil{Mt. Stromlo and Siding Spring Observatories \\
Australian National University, Weston, ACT 2611, Australia \\
E-mail: {\tt robyn@mso.anu.edu.au}}

\author{D.R. Alves}
\affil{Lawrence Livermore National Laboratory, Livermore, CA 94550 \\
E-mail: {\tt alves@beowulf.llnl.gov}}

\author{T.S. Axelrod}
\affil{Mt. Stromlo and Siding Spring Observatories \\
Australian National University, Weston, ACT 2611, Australia \\
E-mail: {\tt tsa@mso.anu.edu.au}}

\author{A.C. Becker}
\affil{Department of Astronomy, University of Washington, Seattle, WA 
98195 \\
E-mail: {\tt becker@astro.washington.edu}}

\author{D.P. Bennett}
\affil{Lawrence Livermore National Laboratory, Livermore, CA 94550 \\
E-mail: {\tt bennett@beowulf.llnl.gov}}

\author{D.F. Bersier}
\affil{Mt. Stromlo and Siding Spring Observatories \\
Australian National University, Weston, ACT 2611, Australia \\
E-mail: {\tt bersier@mso.anu.edu.au}}

\author{K.H. Cook}
\affil{Lawrence Livermore National Laboratory, Livermore, CA 94550 \\
E-mail: {\tt kcook@beowulf.llnl.gov}}


\author{K.C. Freeman}
\affil{Mt. Stromlo and Siding Spring Observatories \\
Australian National University, Weston, ACT 2611, Australia \\
E-mail: {\tt kcf@mso.anu.edu.au}}

\author{K. Griest, J.A. Guern, M. Lehner}
\affil{Department of Physics, University of California, San Diego, CA 
92093 \\
E-mail: {\tt griest@astrophys.ucsd.edu, jguern@astrophys.ucsd.edu,
matt@astrophys.ucsd.edu}}

\author{S.L. Marshall}
\affil{Lawrence Livermore National Laboratory, Livermore, CA 94550 \\
E-mail: {\tt stuart@beowulf.llnl.gov}}

\author{D. Minniti}
\affil{Lawrence Livermore National Laboratory, Livermore, CA 94550\\
and Departmento de Astronomia y Astrofisica, P. Universidad Catolica,
Casilla 104, Santiago 22, Chile \\
E-mail: dminniti@llnl.gov}

\clearpage

\author{B.A. Peterson}
\affil{Mt. Stromlo and Siding Spring Observatories \\
Australian National University, Weston, ACT 2611, Australia \\
E-mail: {\tt peterson@mso.anu.edu.au}}

\author{M.R. Pratt}
\affil{Department of Astronomy, University of Washington, Seattle, WA 
98195 \\
E-mail: {\tt mrp@astro.washington.edu}}

\author{P.J. Quinn}
\affil{Europen Southern Observatory, D-85748 Garching bei M\"unchen, Germany \\
E-mail: {\tt pjq@eso.org}}

\author{A.W. Rodgers}
\affil{Mt. Stromlo and Siding Spring Observatories \\
Australian National University, Weston, ACT 2611, Australia
}

\author{C.W. Stubbs}
\affil{Department of Astronomy, University of Washington, Seattle, WA 
98195 \\
E-mail: {\tt stubbs@astro.washington.edu}}

\author{W. Sutherland}
\affil{Department of physics, University of Oxford, Oxford OX1 3RH, U.K. 
\\
E-mail: {\tt wjs@oxds02.astro.ox.ac.uk}}

\author{A. Tomaney}
\affil{Department of Astronomy, University of Washington, Seattle, WA 
98195 \\
E-mail: {\tt austin@astro.washington.edu}}

\clearpage

\author{T. Vandehei}
\affil{Department of Physics, University of California, San Diego, CA 
92093 \\
E-mail: {\tt vandehei@astrophys.ucsd.edu}}

\author{D.L. Welch}
\affil{Department of Physics and Astronomy, McMaster University, \\
    Hamilton, Ontario L8S 4M1, Canada \\
E-mail: {\tt welch@physics.mcmaster.ca}}

\author{The MACHO Collaboration}


\begin{abstract}

We present an analysis of the period distribution of $\sim 1800$ Cepheids
in the Large Magellanic Cloud, based on data obtained by the MACHO microlensing
experiment and on a previous catalogue by Payne-Gaposchkin. Using stellar evolution
and pulsation models, we construct theoretical period-frequency distributions that are
compared to the observations. These models reveal that a significant burst 
of star formation has occurred recently in the LMC ($\sim 1.15\times 10^8$ years).
We also show that during the last $\sim 10^8$ years, the main center of star
formation has been propagating from SE to NW along the bar. We find that the
evolutionary masses of Cepheids are still smaller than pulsation masses by
$\sim 7$ \% and that the red edge of the Cepheid instability strip could be slightly
bluer than indicated by theory. There are $\sim 600$ Cepheids with periods
below $\sim 2.5$ days cannot be explained by evolution theory. We suggest
that they are anomalous Cepheids; a number of these stars are double-mode Cepheids.

\end{abstract}


\section{Introduction}

The star formation history of the Large Magellanic Cloud has attracted a lot of attention,
particularly during the last couple of years. The global picture that has emerged
is that this galaxy has been forming stars at a low rate during most of its lifetime
and suddenly the star formation rate increased by a large factor about 3 Gyr ago
(Bertelli {\it et al.} 1992, Gallagher {\it et al.} 1996). Much is known also on
the very recent star formation (see Hyland 1991 and Kennicutt 1991 for reviews). 
Between these time scales of $\sim 10$ Myr and $\sim 1$ Gyr, it has been suggested
by several authors that there had been a significant increase of the star formation 
rate (e.g Girardi {\it et al.} 1995, Wood {\it et al.} 1985). This is typically the
age range of Cepheids ($\sim 100$ Myr) and it is exactly
this window that Becker, Iben \& Tuggle (1977, hereafter BIT) have explored.
Using evolution and pulsation models, they have produced theoretical Cepheid
period-frequency distributions.

Several factors determine the shape of the Cepheid period distribution in a galaxy.
Among these are the details of the evolution prior to and during the Cepheid phase,
the location of the instability strip and the metallicity of the galaxy (or the range
in metallicity). The Star Formation Rate (SFR) and the Initial Mass Function (IMF)
are external parameters also controlling the number of Cepheids as a function of
period. While all these parameters are merged into a single period distribution,
they do not all influence it in the same way. By comparing a theoretical
distribution with the observed one, we can thus hope to constrain the ingredients
making up this distribution.
One of the main conclusion of BIT was an ordering in metallicity between the SMC, LMC,
our Galaxy and M31, from low- to high-Z, respectively. They could roughly reproduce
the distribution observed for each galaxy but they had to invoke a ``two-component
birthrate function'' and suggested that the LMC had experienced a burst of star
formation a few tens of millions of years ago. The Cepheids sample they used for the LMC
was taken from Payne-Gaposchkin's catalogue (1971) of 1110 Cepheids, based on Harvard
photographic plates accumulated over $\sim 60$ years.
With a new generation of stellar models incorporating up-to-date input physics
(Fagotto {\it et al.} 1994) and, most importantly, with a larger Cepheid sample obtained
as a by-product of the MACHO microlensing experiment, we thought it was timely to
repeat the exercise, with a more modern and more accurate approach.

\section{Observations and the Cepheid catalog}

\subsection{Observations}

The MACHO project is a microlensing experiment designed to search for dark matter
in the halo of the Milky Way. The observing strategy is to monitor a large number
of stars ($\sim 10^7$) each night in the direction of the Large and Small
Magellanic Clouds. At this time, 82 LMC fields covering approximately 40 square
degrees have been monitored for $\sim 6$ years.

The MACHO project uses a dedicated 1.27m telescope (located at Mt Stromlo, Australia).
A specially designed instrumental system consists of wide-field corrective optics,
a dichroic beam splitter and wide-pass ``red'' and ``blue'' filters.  Each focal
plane contains a mosaic of four 2048$\times$2048 Loral CCDs. Imaging at the prime
focus gives a field of view of 0.5 square degree. For a complete description of
the telescope and camera systems, see Hart {\it et al.} (1996) and Stubbs
{\it et al.} (1993) respectively. Profile-fitting photometry is performed in
real-time with {\sc SODoPHOT}, derived from {\sc DoPHOT} (Schechter, Mateo \& Saha
1994). The microlensing analysis (and its byproduct, variable star catalogs)
is accomplished with subsets of the LMC data thus far obtained.
The catalog of Cepheids analysed in this paper have been identified
in the data from 22 fields used for the ``two-year'' LMC microlensing analysis 
(Alcock {\it et al.} 1997a), to which the reader is referred for further details
(i.e. the fields observed, total number of images reduced).

\subsection{The period distribution}

Specific searches for Cepheids in the Macho database have been
conducted yielding a catalog of $\sim 1700$ variables.  The search
procedure will be described in detail elsewhere (Alcock et al 1998).
For our purposes, we primarily require the accurate measurements of
the pulsation periods and identifications of pulsation mode (i.e.
fundamental or first overtone). A number of stars (63) appear
twice in this catalogue as they are in two overlapping fields; for each of these
one entry has been removed from the list. We then added 72 double-mode Cepheids
that have been explicitly searched for (see Alcock
et al. 1995b for the search procedure and Welch {\it et al.} 1997 for the
latest sample).

\begin{figure}[htb]
\epsfxsize=14cm \centering{\mbox{\epsfbox{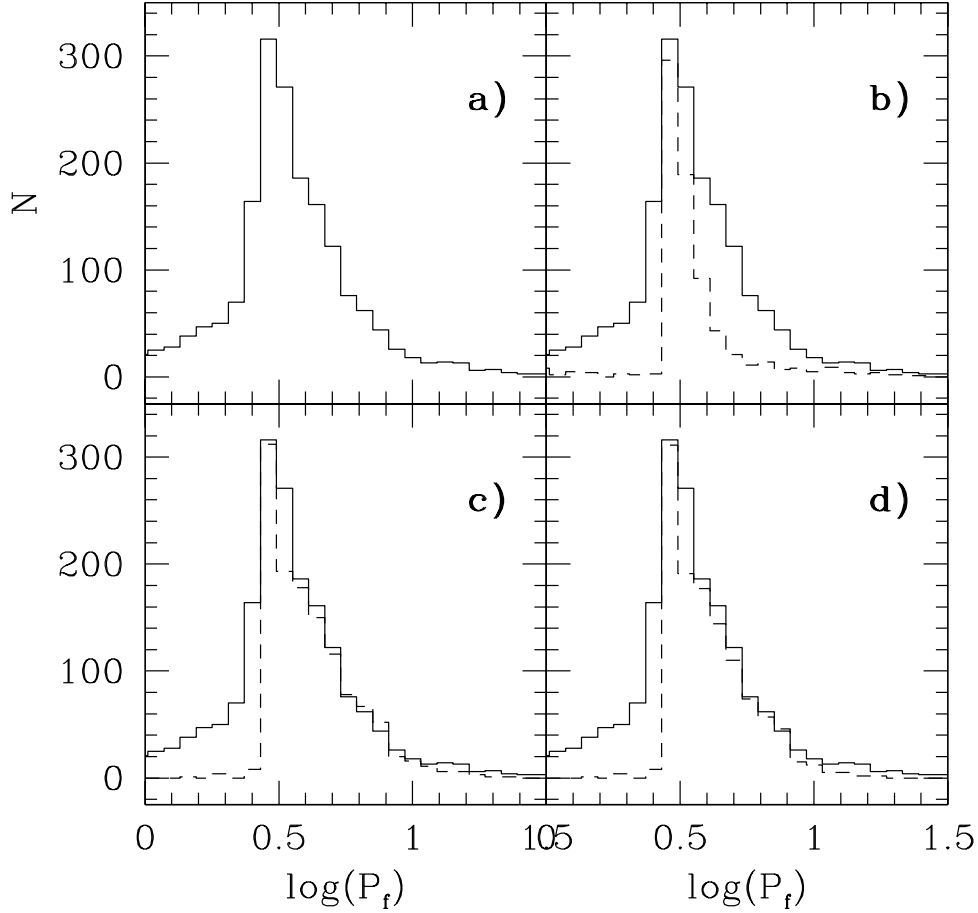}}}
\caption[ ]{{\bf a)} The final period distribution, after
transformation of the first overtone periods into fundamental periods.
{\bf b)} Assuming a constant star formation rate over time and a slope of $-2.35$ for
the IMF gives the distribution shown as a dashed line.
{\bf c)} An increase in the star formation rate of a factor of $\sim 3$ centred on
$\sim 110$ Myr allows to reproduce the long-period side of the distribution
{\bf d)} This model shows the effect of a change of slope of the red edge,
the strip being narrower at lower luminosities. This creates fewer short-period
Cepheids. The burst at $1.1\,10^8$ Myr is thus less intense}
\label{fig_obs_models}
\end{figure}

We have cross-correlated our sample with another major catalog of LMC Cepheids
published by Payne-Gaposchkin (1971, hereafter referred to as PG). The total
number of PG Cepheids in our 22 fields is 551. For each PG star, we searched
a corresponding MACHO Cepheid, requiring a match between coordinates and
period. Our search for Cepheids has been iterative in nature and at each step,
the number of PG stars not matching any MACHO star has been decreasing.
The present sample contains only 70 PG Cepheids, not observed by MACHO
(see Fig.~\ref{fig_obs_models}).

\begin{figure}[htb]
\epsfxsize=14cm \centering{\mbox{\epsfbox{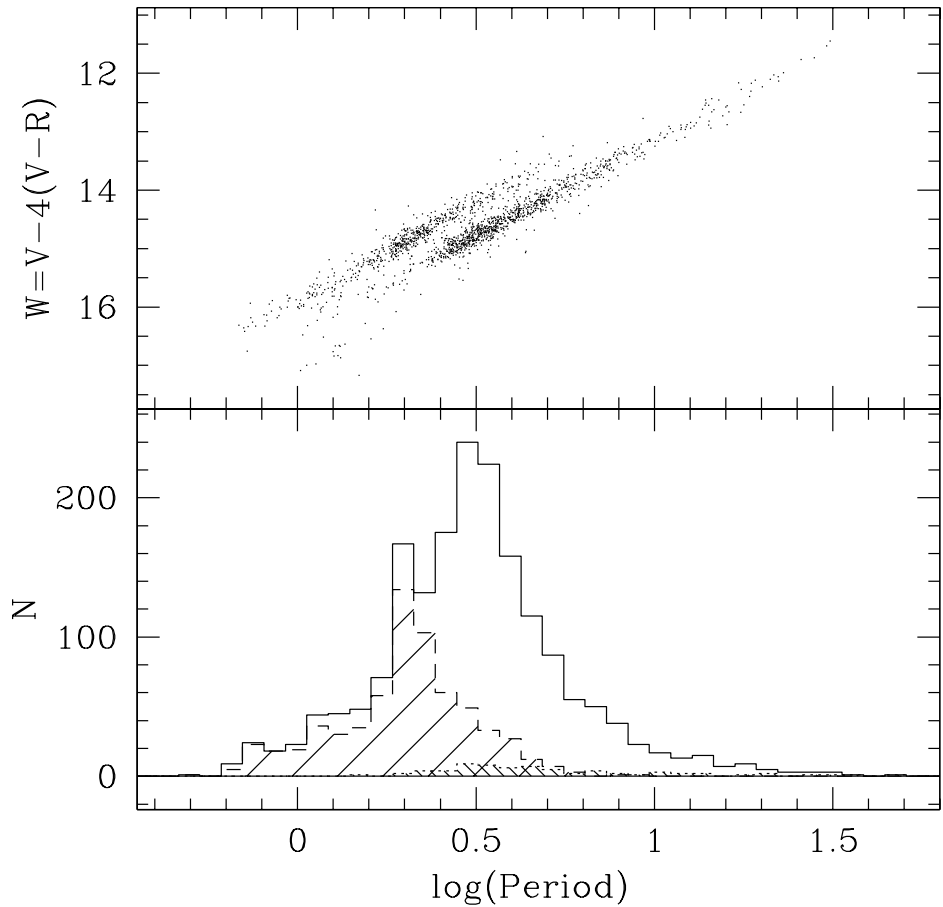}}}
\caption[]{ {\bf a)} The quantity $W$ is defined as $W=V-4(V-R_C)$.
It is essentially
reddening-free. This allows to clearly separate fundamental and first overtone
pulsators. These two sequences are obvious in this figure. Note that the PG
stars are not plotted, by lack of color information. The tightness of this relation
and the absence of bright outliers seem to rule out the existence of a significant
foreground population (Alcock {\it et al.} 1997b). {\bf b)} The observed
period-frequency distribution. The dense shading illustrates the PG contribution,
the wider shading is for overtone pulsators.
\label{fig_per_obs} }
\end{figure}

As will be explained in more detail later, the periods of Cepheids coming from
our population synthesis models are for the fundamental mode. In order to compare
the same things, we have transformed the observed first-overtone periods into fundamental
periods. The mode identification is based on the PL relation (Fig.~\ref{fig_per_obs}).
In a few cases, we also used the parameters $R_{21}$ and $\phi_{21}$
(Simon \& Lee 1981) resulting from a twelfth-order Fourier decomposition.
The result is illustrated in Fig.~\ref{fig_per_obs}. This allows to transform
the $1^{st}$ overtone periods into fundamental periods since the ratio of
first-overtone to fundamental mode period is known from double mode Cepheids
(Alcock {\it et al.} 1995):
\begin{displaymath}
P_1/P_0 = 0.733-0.034\,\log P_1
\end{displaymath}
where $P_0$ and $P_1$ are the fundamental and the first-overtone periods respectively.
This separation of the pulsation modes represents a major improvement over the
work of BIT since they were using the PG sample
that is deficient in short-period Cepheids compared to the MACHO sample.
These stars are particularly difficult to detect when they are close to
the detection limit of photographic plates. Furthermore our CCD photometry gives
excellent quality mean magnitudes, as illustrated in Fig~\ref{fig_per_obs}. A
period-magnitude plot for the PG sample does not reveal two clearly separated
sequences. This transformation modifies the period-frequency distribution
significantly. The result is shown on Fig.~\ref{fig_obs_models}.
The period-frequency distribution is characterised by a pronounced maximum
at $\log P \simeq 0.48$, a sharp break at $\log P\simeq 0.4$, a long period tail after
the maximum and a short period tail, at periods shorter than the break
($\sim 600$ stars).
Table~\ref{tab_periods} summarises this in numerical form.

%
%

\begin{table}
\dummytable\label{tab_periods}
\end{table}

In this paper we make the critical assumption that the Cepheid
catalog has no systematic incompleteness as a function of period.
Incompleteness almost certainly results from the reduced effective 
area of the fields monitored due to details of the instrument and
data reduction procedure of MACHO.
There is a number of arguments that allow us to consider that we have a
fair sample, not affected by any bias. The maximum of the MACHO
distribution is at the same period as for the whole PG sample (551 stars in our 22
fields). Given the important differences in observational set-up and techniques
used to search for variable stars, we think this maximum is a true feature of
the LMC Cepheid population.
The break at $\log P \simeq 0.4$ is so sharp that it cannot be caused by
a sudden decrease in variable star detection efficiency. Furthermore the
detection of $\sim 8000$ RR Lyrae, that are 2 to 3 magnitudes fainter than Cepheids at the
break, strongly suggests that the Cepheid catalog is \emph{not} magnitude
limited (and thus biased toward incompleteness at short periods).
Since the PG Cepheids not found in
the MACHO catalog represent only 4\% of the total number, we have
included them in our analysis. This mixing of catalogs has a negligible
effect on the results of this paper.

With the arguments given above, we are confident that our sample is
\emph{representative} of the true
distribution of Cepheids. As a consequence, we will require from our models
to reproduce the global shape of the period distribution, we will not put
too much weight on reproducing each feature in great detail.

\section{The star formation history}

\subsection{Construction of the models}

To produce a theoretical period distribution, one needs evolutionary tracks and
pulsation models. The tracks will tell when a star of a given mass will be at a
given position in the HR diagram and the pulsation models will give the position
of the instability strip and the period of the stars that are within the strip.
We have used evolutionary models computed by Fagotto {\it et al.} (1994) at a
metallicity $Z=0.008$ corresponding to the mean metal content of young stars in the LMC.
The pulsation models are from Capitanio {\it et al.} (in preparation). They are
linear models, computed in exactly the same way as in Chiosi {\it et al.} (1993)
but using OPAL opacity data. For the sake of clarity, we give below the edges
of the instability strip resulting from these models and the relation
allowing to compute the period from $M,\ L$ and \teff. These equations are
\begin{eqnarray}
\log(\teff) & = & 3.9281+0.0302\cdot\log (M) \\
            &   & -0.0497\cdot\log (L) -0.0003\cdot(\log (L))^2 \nonumber
\end{eqnarray}
for the fundamental blue edge,
\begin{eqnarray}
\log(\teff) & = & 3.8902+0.0716\cdot\log (M) \\
            &   & -0.044\cdot\log (L)-0.0055\cdot(\log (L))^2 \nonumber
\end{eqnarray}
for the fundamental red edge,
\begin{eqnarray}
\log(\teff) & = & 3.7644+0.1135\cdot\log (M) \\
            &   & -0.0647\cdot\log (L)-0.0239\cdot(\log (L))^2 \nonumber
\end{eqnarray}
for the first overtone blue edge and
\begin{eqnarray}
\log (P) & = & 12.2229+0.6886\cdot\log (L) \\
         &   & +0.0245\cdot(\log (L))^2-0.5862\cdot\log (M) \nonumber \\
         &   & -3.3110\cdot\log (T) \nonumber
\end{eqnarray}
for the $P,\ M,\ L,\ \teff$ equation (fundamental mode).

A model of the star formation history is constructed by assigning a mass to a number
of ``synthetic'' stars, the constraint being that this sample obey a specified Initial
Mass Function (IMF), then distributing these stars in time according to some given
star formation rate (e.g. with a constant star formation rate or a Gaussian burst).
Since evolutionary tracks are given for a finite set of masses, the next step is
an interpolation in $\log L$ and \teff\ in the tracks to get these parameters for
each synthetic star. A test is then performed to determine the position of each star
with respect to the instability strip; if the star is in the strip, the period is
computed.
This is done until the number of synthetic Cepheids is approximately the same as
the observed number. The IMF has been assumed to be a power law with an exponent
$x=-2.35$  (Salpeter 1955).

A complication arises from the fact that there is a region of the instability strip
where, according to theory, both fundamental and $1^{st}$ overtone pulsators can
exist.
However we do not know at present how a given star in this overlapping region
``chooses'' to pulsate in one mode or the other. We thus decided to assign a
fundamental period to \emph{all} stars in the strip. Only for stars outside
the fundamental instability strip but inside the first overtone strip the period
has been computed according to the relation for the first overtone. This lack of
knowledge about the mode selection is the reason why, in the observed distribution,
all overtone periods have been transformed into fundamental periods.

\subsection{The effect of evolution models}

The first and simplest model that we can build is with a constant star formation rate
and an IMF slope of $-2.35$. It is displayed on Fig.~\ref{fig_obs_models}b.
While the break is relatively well reproduced, the maximum appears too early. There is
also a deficit of long-period Cepheids and we cannot reproduce the short-period
part of the distribution. We first concentrate on the position of the maximum.
To obtain a maximum at a longer period, we can
increase the luminosities of the model Cepheids or reduce the masses. An increase in
luminosity actually shifts the whole distribution to larger periods, an undesirable
result since the break is no longer reproduced correctly. It turns out that a
decrease of the masses by $\sim 7$\% allows to reproduce the maximum \emph{and} the
break simultaneously. The position of the break corresponds to the mass at
which the blue loop is not extended enough to reach the instability strip. Below
this mass (which is $\sim 3.8\, M_\odot$ -- modified mass) stars do not become
Cepheids.

A decrease in the mass means that, for a given initial mass, the luminosity is
increased but also the main-sequence lifetime is increased. This is why a change of
the masses does not affect the period distribution in the same way as a change of the
luminosities, since the moment when a star will become a Cepheid is also changed.
A physical interpretation of this effect is that more overshooting is needed during
core hydrogen-burning. To first order this has the same effect as our adopted
modification of the masses.
This 7\% decrease in the masses could also be caused by mass loss. However this
is very unlikely because it would imply mass-loss rates much larger than what are
actually observed for intermediate mass stars on the red-giant branch (de Jager
{\it et al.} 1988).

Talon {\it et al.} (1997) recently showed that
a large rotation velocity ($\sim 300$ km/s) significantly increases the luminosity
at the end of the main sequence phase. Since their $9\,M_\odot$ model has not been
evolved beyond the end of core hydrogen burning, it would be premature to claim that
rotation should be preferred over any other mechanism but at least it points in
the right direction. 

A problem still remains with the short- and long-period
tails. A natural way of producing more long-period Cepheids is to create more stars,
that is to increase the star formation rate at the right time.
A Gaussian increase, centred at $\sim 1.1\times 10^8$ years, with a width (sigma) of
$2.4\times 10^7$ years gives a relatively
good agreement (see Fig.~\ref{fig_obs_models}). At the maximum, this represents an SFR
increase of a factor of about 7.
This increase of the SFR comes as no surprise since it can actually be seen in the
HR diagram (e.g. Sebo \& Wood 1995). It creates a clump of yellow-red supergiants
at $V-I \simeq 1$ mag.

For all the models that we have constructed, we always had slightly too large a number
of Cepheids around the maximum. It was not possible to have simultaneously the
right number of stars at the peak of the distribution and the correct number beyond
5 days. This could be solved by invoking the existence of another component
in the star formation (e.g. a very recent, low-amplitude burst). It could also
reveal a problem with the evolutionary tracks. For instance, a change in the extension
of the blue loops would impact the period distribution. Another possibility could be
a problem with the core-helium burning lifetime of $\sim 4.5 M_\odot$ stars.
This would also change the amplitude of the burst, in the sense that it would be lower
if the Cepheid life-times were shorter for $\sim 4.5 M_\odot$ stars.

%
%

\subsection{Morphological distribution of Cepheids}

\begin{figure}[htb]
\epsfxsize=14cm \centering{\mbox{\epsfbox{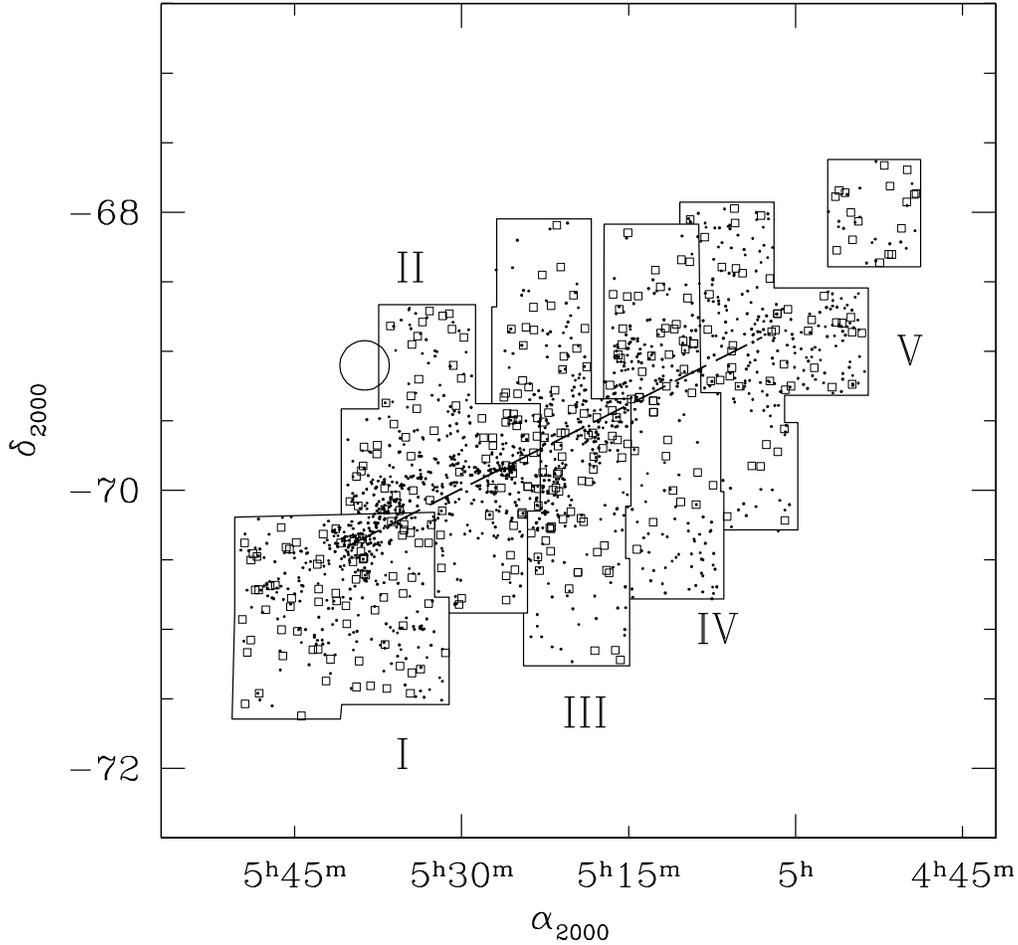}}}
\caption[ ]{Our 22 fields have been separated into five regions, going from
south to north, roughly along the bar. The thick dashed line outlines the bar,
the circle indicates the position of 30 Dor. The small dots are for Cepheids
with $\log\, P\geq 0.4$ while open squares are for Cepheids with $\log\, P< 0.4$.
Note the concentration of classical ($\log\, P\geq 0.4$) Cepheids in the bar}
\label{fig_regions}
\end{figure}

As mentioned above, there is a clear pattern in the distribution of Cepheids as a
function of period (e.g. Payne-Gaposchkin 1974). To explore this more quantitatively,
we divided our sample into five regions, going from south to north of the bar.
Figure~\ref{fig_regions} shows these regions with the bar outlined.
%
%

\begin{figure}[htb]
\epsfxsize=14cm \centering{\mbox{\epsfbox{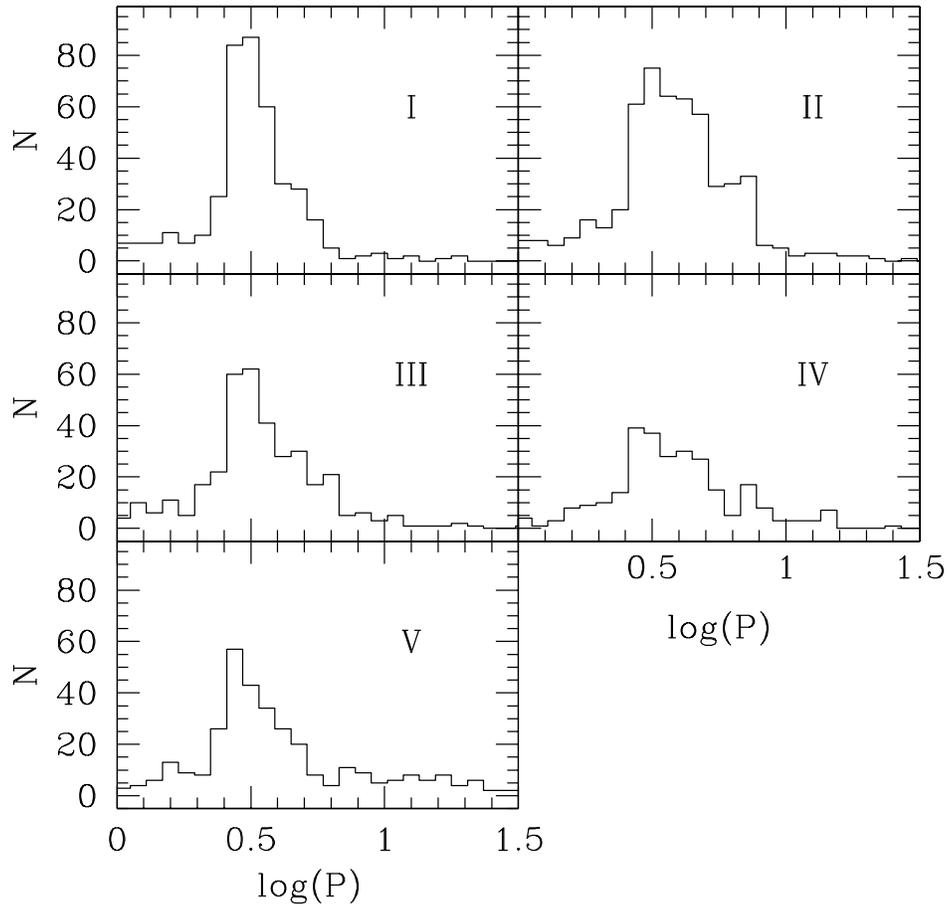}}}
\caption[ ]{The period-frequency distribution for the five regions described
in the text.}
\label{fig_regdist}
\end{figure}

It can be seen on figure~\ref{fig_regdist} that the break in the distribution is at
the same period for each of the five regions. This break being a very sensitive
function of metal abundance, we conclude that there is no mean abundance change
across the bar of the LMC.
%
%
For each region we have computed a model of the star formation history following
the same procedure as above. The aim was also to determine the center (in time)
of the burst and its duration ($\tau_b$ and $\sigma_b$ respectively).
Our main result is a clear displacement in time
of the ``star formation centre'' from regions I to region V. The position of the
Gaussian burst appears earlier and earlier as we move from the SE end of the bar
to the NW end (see Table~\ref{tab_reg}). The duration of the burst in each region
is roughly constant. For each region, we estimated the uncertainty on $\tau_b$ to
be $\sim 2 $ Myr.

\begin{table}
\dummytable\label{tab_reg}
\end{table}

%


With the numbers given in Table~\ref{tab_reg}, we can compute the velocity
at which the star formation peak has been propagating in the bar (assuming
a distance of 50 kpc for the LMC). It took $40\,10^6$ yr to go from region I
to region V which represents a velocity of $\sim 100$ km/s.
This large velocity gives support to the fact that star formation has been
tidally triggered during the last passage of the LMC close to the Milky Way
which, according to Lin {\it et al.} (1995), happened roughly 150 Myr ago.

\section{Tests of the pulsation models}

Up to now, we have only derived constraints for the evolution models. However,
it is obvious that pulsation
also influences the period distribution through the position of the edges of the
instability strip. We have performed a series of tests of the pulsation models by
shifting the blue and/or red edge in temperature. Small changes (of order 0.01 in
$\log T_e$) could be accommodated by modifying the IMF or the SFR. Changes larger
than 0.02 do not give a good agreement with the observed distribution.
We summarise these results as follows:
\begin{itemize}
  \item Making the blue edge bluer: the distribution is too wide, with an excess of
	stars around the maximum. This also produces far too many overtones.
  \item Making the blue edge redder: there is a deficiency of Cepheids between
	$0.5\leq \log P \leq 0.75$
  \item Making the red edge bluer: there is a significant lack of long-period Cepheids
	that can not be compensated for by a change in other parameters
  \item Making the red edge redder: there are too many stars around the maximum.
  \item Making both edges bluer: the maximum is flat, not peaked. Furthermore there
	are too many long-periods Cepheids.
  \item Making both edges redder: this gives too many Cepheids at the maximum,
	the effect is almost the same as making the red edge redder.
\end{itemize}
In other words, a widening of the instability strip produces too many Cepheids
around the maximum and it is impossible to obtain a satisfying agreement even when
changing the IMF and SFR. A change in one of the edges at a time changes the skewness
of the distribution.

Quantitatively, the position of the blue edge is good to $\pm 0.01$ in
$\log T_{\mathrm eff}$. The position of the red edge is less well constrained
and $\pm 0.02$ is a reasonable estimate of the uncertainty. Changes in the IMF and
SFR, simultaneously with the position of the red edge, can give a satisfactory
agreement, and compensate for the change in the red edge. The overall
shape and the details of the distribution are more sensitive to the position of the
blue edge than to that of the red edge.

A change in the slope of the red edge, in the
sense that makes the instability strip narrower at low luminosities, gave a
good result. The resulting distribution is shown on Fig.~\ref{fig_obs_models}d.
A narrower strip at low luminosities will produce fewer short-period Cepheids.
It is therefore easier to create the large number of Cepheids that a constant
SFR model can not reproduce. This is why the amplitude of the Gaussian component
of the SFR is less important, but still necessary. With a modified red edge,
the model gives a marginally better agreement with the data. It does not
change significantly any of our conclusions regarding the star formation history.
Therefore we decided to use the original instability strip.

\section{The short-period Cepheids}

A major failure of the models is that it is impossible to create Cepheids
with periods smaller than $\sim 2.5$ days ($\log (P) \simeq 0.4$).
The reason is that the blue loops for $\sim 3\,M_\odot$ stars do not reach
the instability strip. One could imagine changing the evolutionary path
of these objects by extending their blue loops. For intermediate-mass stars,
most of the core helium burning phase is spent in a band that
extends from the upper left part of the HR diagram to the lower right part. To explain
the presence of these short-period Cepheids in a classical framework, this
band should bend back towards higher temperatures, which means that going to lower and
lower masses, the blue loops should be more and more extended. This is unlikely
given our knowledge of stellar evolution. Furthermore, since the bluest part of
the blue loops are actually where the stars spend most of the core helium burning
time, stars on this bent sequence should be observed in open clusters. The fact
that cluster HR diagrams show no sign of such a sequence rules out the possibility
of extended blue loops for stars with masses lower than $3.5\, M_\odot$.

Another possibility could be a biased IMF, producing an extremely large
number of relatively low mass stars. Models were run with an IMF biased toward low
mass stars, however without success. The reason is actually the same as for
the break: below a given mass, stars do not become Cepheids. All
stars with periods lower than this limit cannot be considered as classical
Cepheids (young intermediate-mass stars in the core helium burning phase).

We suggest that these stars are anomalous Cepheids, formed through the merging of a
binary star system. This scenario
of binary coalescence is often mentioned as the formation mechanism of anomalous
Cepheids and blue stragglers (although direct stellar collisions may be a more
likely mechanism for blue stragglers in dense globular clusters). 
A recent review of the properties of anomalous Cepheids (ACs) can be found in Nemec
{\it et al.} (1994) as well as a complete census of known anomalous Cepheids and we will
briefly recall the most important facts. They have been found in dwarf spheroidal
galaxies and they are $\sim 1-2$ magnitudes brighter than RR Lyrae.
They are believed to be formed through the merging of two stars close to the
turn-off of old stellar systems (the progenitors having masses of
$\sim 0.7-0.8\,M_\odot$). Evidence for this comes from the high pulsation
masses found for anomalous Cepheids which are estimated to be $\sim 1.5\, M_\odot$
(e.g. Wallerstein \& Cox 1984).

We computed the masses of these stars in the following way. We transformed our
observed $V-R$ data\footnote{Comparison of our data with published
standard star sequences in the LMC shows that our photometric accuracy in
magnitude and color is 0.035 mag.} to $V-I$ using data in Bessell {\it et al.}
(1998). Equations~(22) and (23) of Chiosi {\it et al.} (1993) give the masses
as a function of $M_V,\ (V-I)$ and $\log (P)$ (similar equations for the
new pulsation models were not available to us), where we have used a
distance modulus of 18.5 and a constant reddening $E(B-V)=0.1$.
The distribution of masses is peaked at 3.2$\, M_\odot$, with 61\% of the
stars having a mass between 2.5 and 3.5$\,M_\odot$. Furthermore,
they are almost all brighter than $V=17$, which means that they would be very bright
and massive ACs. The possible solution actually resides in the history of
the LMC. Evidence has been accumulating in the last few years that the LMC
is a young galaxy, by astronomical standards. Most of the stars have been
formed in a burst $\approx 3-4$ Gyr ago (e.g. Bertelli {\it et al.} 1992,
Gallagher {\it et al.} 1996), with a possible peak in the SFR $\sim 2$ Gyr ago.
A 2 Gyr old star at the turn-off, with a metallicity [Fe/H]$=-0.6$ has a mass of
$\sim 1.4 \,M_\odot$. The merging of two such stars would produce an object with
a mass in the observed mass range, it would also be more luminous. Having such
a mass and metallicity, a merged binary could easily reach the instability
strip and thus become an anomalous Cepheid.

Another piece of evidence in favour of this scenario is given by double-mode Cepheids,
since a large number of these have (fundamental) periods shorter than $\sim 2.5$
days (the period of the break). Masses can be obtained for double mode Cepheids
from pulsation theory. Morgan \& Welch (1997) have determined possible masses
for double-mode Cepheids and found fairly low values. One of their main
conclusions is that double-mode Cepheids with $\log P\leq 0.4$
are on a non standard mass-luminosity relation, namely that determined by Simon (1990)
that makes stars very luminous for a given mass.
The masses and luminosities of the short-period Cepheids indicate that the
mass-luminosity relation valid for these stars is close to that of Simon.
This is in excellent agreement with the short-period double-mode Cepheids.

Figure~\ref{fig_regions} shows the morphological distribution of LMC Cepheids.
The visual impression is that Cepheids with $\log\, P \geq 0.4$ are more
concentrated in the bar than short-period Cepheids. To test this idea further,
we performed a two-dimensional Kolmogorov-Smirnov test on the positions
in order to determine whether these two groups of Cepheids could originate
from the same population. The result of this test is that we can exclude at
the 99\% confidence level the fact that these two groups of Cepheids originates
from the same parent population. We also performed tests with different period
sub-samples (e.g. $0\leq\log\, P\leq 0.4$ compared to $0.5\leq\log\,P \leq 0.6$)
and these consistently gave the same result.
This shows that classical Cepheids are sufficiently young
that they have not yet been dynamically mixed, whereas ACs must have been mixed.
This is another argument in favour of an intermediate-old origin of these objects.


\section{Conclusions}

The strongest conclusion is that the evolution models have to be modified in a sense
that would correspond to a $\sim 7\%$ decrease in mass. This means that for a given
ZAMS mass, the luminosity will be larger and the core hydrogen burning time will be
longer. Both effects would be natural consequences of an increase of the overshoot
parameter $\lambda_c$. However increasing the overshoot could influence the extension
of the blue loops, in the sense that they would be shorter (see Alongi {\it et al.}
1991). Another possible solution could be rotation (Talon et al. 1997).
Models for intermediate mass stars, including rotation,
computed up to the end of helium burning would be extremely interesting in this
context.

Modifying the evolutionary tracks, the position of the strip, the slope of the IMF
or the SFR produce different results on the theoretical distribution. While there
is some kind of ``degeneracy'' (for instance a small change in the position of the red
edge can be partially compensated by a modification of the SFR), we can derive
valuable
quantitative constraints on stellar models and on the star formation history.
Whereas Becker {\it et al.} (1977) obtained a qualitative agreement between the
observed and theoretical distributions for the LMC, the present sample allows to
infer quantitative constraints on the stellar models. The differences between
the present study and BIT come from the use of improved stellar models, based
on up-to-date input physics, and from a much better observational sample that
allows clear mode identification. While an increase of star formation at $\sim 100$
Myr had already been put forward by several authors, we have been able to quantify
this more precisely and to derive a time scale for the propagation of star formation
along the bar. This might help to improve dynamical models of the LMC-SMC-Milky Way
system.

The IMF does not have a strong influence on the results because changing the IMF can
be compensated for by modifying the SFR. However, we can can rule out a very large
change of the slope of the initial mass function. The models allow us to constrain
the slope to $2.35 \pm\sim 0.6$. The fact that our results do not depend strongly
on the IMF makes our conclusions on the SFR more robust.

It has been shown that, while stellar models can account for the bulk of the
Cepheids observed in the LMC, there is a significant number of stars, all of
fundamental period shorter than $\sim 2.5$ days, that can not be considered as
classical Cepheids.
The fact that these stars could be the end product of a binary coalescence is an
appealing solution. It would resolve the mass discrepancy problem observed for
double-mode Cepheids and provide a natural explanation for the relatively low masses
of these stars. Given its star formation history in the last few Gyr, the LMC provides
a reservoir of potential progenitors of Anomalous Cepheids. The stars we have
found may be the \emph{massive, luminous and metal-rich} analogs of anomalous
Cepheids found in dwarf spheroidal galaxies.

\acknowledgements

Work performed at Lawrence Livermore National Laboratory (LLNL)
is supported by the Department of Energy (DOE) under contract 
W7405-ENG-48. Work performed by the Center for Particle Astrophysics 
(CfPA) on the University of California campuses is supported in part 
by the Office of Science and Technology Centers of the National 
Science Foundation (NSF) under cooperative agreement AST-8809616. 
Work performed at MSO is supported by the Bilateral Science and
Technology Program of the Australian Department of Industry,
Technology and Regional Development. DFB gratefully acknowledges
support from the Swiss National Fund for Scientific Research.
KG acknowledges a DOE OJI grant, and the support of the Sloan
Foundation. DLW was a Natural Sciences and Engineering Research
Council (NSERC) University Research Fellow during this work.
It is a pleasure to thank Peter Wood for numerous enlightening
discussions and for allowing us to use his results in advance
of publication.

\newpage

\begin{deluxetable}{ c c c c c c c c c c }
\footnotesize
\tablenum{1}
\tablecaption{This table gives the period-frequency distribution in
numerical form.}
\label{tab_periods}
\tablewidth{0pt}
\tablehead{
\colhead{Bin\tablenotemark{a}} & \colhead{Total\tablenotemark{b}} &
\colhead{F\tablenotemark{c}} & \colhead{O\tablenotemark{d}} & \colhead{PG\tablenotemark{e}} &
\colhead{Bin\tablenotemark{a}} & \colhead{Total\tablenotemark{b}} &
\colhead{F\tablenotemark{c}} & \colhead{O\tablenotemark{d}} & \colhead{PG\tablenotemark{e}} }
\startdata
$ -0.22\ -0.16$ &   0 &   0 &   5 &  0 &  0.80   0.86   &  58 &  49 &   0 &  5 \nl
$ -0.16\ -0.10$ &   1 &   0 &  22 &  0 &  0.86   0.92   &  44 &  39 &   1 &  2 \nl
$ -0.10\ -0.04$ &   3 &   0 &  18 &  0 &  0.92   0.98   &  22 &  20 &   1 &  1 \nl
$ -0.04\  0.02$ &  24 &   2 &  19 &  0 &  0.98   1.04   &  18 &  17 &   0 &  3 \nl
  0.02   0.08   &  23 &   8 &  34 &  0 &  1.04   1.10   &  12 &  12 &   0 &  1 \nl
  0.08   0.14   &  31 &  15 &  33 &  0 &  1.10   1.16   &  18 &  17 &   0 &  3 \nl
  0.14   0.20   &  46 &  11 &  30 &  1 &  1.16   1.22   &   7 &   7 &   0 &  0 \nl
  0.20   0.26   &  39 &  12 &  55 &  0 &  1.22   1.28   &   7 &   7 &   0 &  0 \nl
  0.26   0.32   &  59 &  26 & 133 &  2 &  1.28   1.34   &   7 &   7 &   0 &  3 \nl
  0.32   0.38   &  75 &  26 & 108 &  3 &  1.34   1.40   &   3 &   3 &   0 &  0 \nl
  0.38   0.44   & 201 & 103 &  60 &  5 &  1.40   1.46   &   3 &   3 &   0 &  1 \nl
  0.44   0.50   & 306 & 185 &  51 &  9 &  1.46   1.52   &   3 &   3 &   0 &  1 \nl
  0.50   0.56   & 265 & 191 &  35 &  8 &  1.52   1.58   &   1 &   1 &   0 &  1 \nl
  0.56   0.62   & 184 & 132 &  24 &  6 &  1.58   1.64   &   0 &   0 &   0 &  0 \nl
  0.62   0.68   & 141 & 100 &  15 &  5 &  1.64   1.70   &   1 &   1 &   0 &  1 \nl
  0.68   0.74   & 121 &  93 &   7 &  7 &  2.06   2.12   &   1 &   1 &   0 &  1 \nl
  0.74   0.80   &  70 &  48 &   3 &  1 &
\enddata
\tablenotetext{a}{interval in $\log (P_f)$}
\tablenotetext{b}{total number in each period interval}
\tablenotetext{c}{number of fundamental pulsators}
\tablenotetext{d}{number of overtone pulsators}
\tablenotetext{e}{number of stars from the PG sample}
\end{deluxetable}

\newpage

\begin{deluxetable}{ c c c c }
\footnotesize
\tablenum{2}
\tablecaption{The position and width of the star formation ``burst'' for each
region.}
\label{tab_reg}
\tablewidth{0pt}
\tablehead{
\colhead{Region} & \colhead{Number} & \colhead{$\tau_b\ (10^8$ years)} & \colhead{$\sigma_b\ (10^7$ years)} }
\startdata
I   & 398 & 1.20 & 2.4 \nl
II  & 518 & 1.13 & 2.2 \nl
III & 365 & 1.08 & 2.4 \nl
IV  & 276 & 1.03 & 2.4 \nl
V   & 332 & 0.80 & 2.8 \nl
\enddata
\end{deluxetable}

\newpage

\end{document}